\documentclass[lettersize,journal]{IEEEtran}
\usepackage{amsmath,amsfonts}
\usepackage{algorithmic}
\usepackage{algorithm}
\usepackage{array}
\usepackage[caption=false,font=normalsize,labelfont=sf,textfont=sf]{subfig}
\usepackage{textcomp}
\usepackage{stfloats}
\usepackage{url}
\usepackage{verbatim}
\usepackage{graphicx}
\usepackage{cite}
\usepackage{amsmath}
\usepackage{amssymb}
\usepackage{bm}
\usepackage{color}
\allowdisplaybreaks[4]
\begin{document}

\title{\huge{Intelligent Reflecting Surface-Aided Electromagnetic Stealth\\over Extended Regions}\vspace{-0.1cm}
}

\author{Qingjie Wu, Beixiong Zheng,~\IEEEmembership{Senior Member,~IEEE}, Guangchi Zhang, \\Derrick Wing Kwan Ng,~\IEEEmembership{Fellow,~IEEE}, and A. Lee Swindlehurst,~\IEEEmembership{Fellow,~IEEE}\vspace{-0.2cm}
\thanks{The work of Beixiong Zheng was supported in part by the National Natural Science Foundation of China under Grant 62201214, Grant 62331022, and Grant 62301171, the Natural Science Foundation of Guangdong Province under Grant 2023A1515011753, Grant 2024A1515010013, and Grant 2022A1515110484, the Fundamental Research Funds for the Central Universities under Grant 2024ZYGXZR087, and the GJYC program of Guangzhou under Grant 2024D03J0006. The work of Guangchi Zhang was supported in part by the Guangdong Basic and Applied Basic Research Foundation under Grant 2023A1515011980, and in part by the Jiangxi Institute of Civil Military Integration Beidou+ Project under Grant 2024JXRH0Y02. \textit{(Corresponding author: Beixiong Zheng.)}}
\thanks{Qingjie Wu and Beixiong Zheng are with the School of Microelectronics, South China University of Technology, Guangzhou 511442, China (e-mail: miqjwu@mail.scut.edu.cn; bxzheng@scut.edu.cn).}
\thanks{Guangchi Zhang is with the School of Information Engineering, Guangdong University of Technology, Guangzhou 510006, China (e-mail: gczhang@gdut.edu.cn).}
\thanks{Derrick Wing Kwan Ng is with the School of Electrical Engineering and Telecom
munications, University of New South Wales, NSW 2052, Australia (e-mail: w.k.ng@unsw.edu.au).}
\thanks{A. Lee Swindlehurst is with the Center for Pervasive Communications
and Computing, University of California, Irvine, CA 92697 USA (e-mail:
swindle@uci.edu).}
}

\maketitle

\begin{abstract}
Compared to traditional electromagnetic stealth (ES) materials, which are effective only within specific frequencies and orientations, intelligent reflecting surface (IRS) technology introduces a novel paradigm for achieving dynamic and adaptive ES by adapting its reflection pattern in real time to neutralize radar probing signals echoed back from the target. In this letter, we study an IRS-aided ES system mounted on an aerial target to evade radar detection admist uncertain/moving radar positions over an extended area. Specifically, we aim to optimize the IRS's passive reflection to minimize the maximum received signal-to-noise ratio (SNR) of the target echo signal in the area. A semi-closed-form solution is derived by first discretizing the continuous spatial frequency deviation to approximate the semi-infinite reflection gain constraint and then leveraging the Lagrange dual method. Simulation results are provided to validate that the proposed IRS-aided ES strategy can consistently reduce the reflection gains for radars located across a large region.
\end{abstract}

\begin{IEEEkeywords}
Electromagnetic stealth (ES), intelligent reflecting surface (IRS), low probability of detection, Lagrange dual method.
\end{IEEEkeywords}

\section{Introduction}
To avoid detection by adversarial radars, electromagnetic stealth (ES) technology is pivotal in attenuating the echo signal power of a radar target, thereby decreasing its detection probability~\cite{ref_Stealth}. Traditional ES relies on coated ES materials that can either redirect incident radar waves towards other non-detective directions or absorb part of the electromagnetic waves at the target surface to reduce the effective radar cross section (RCS)~\cite{ref_Mater_ref,ref_Mater_abs}. However, these ES materials are fabricated to achieve high absorbing efficiency only for specific angles and frequencies of the incident waves, and thus they lack adaptability and flexibility against dynamically varying wireless environments and increasingly advanced radar detection technologies.

Due to its unique capability of proactively manipulating the radio propagation environment in a cost-effective manner, intelligent reflecting surface (IRS) technology has recently emerged as a promising tool for enhancing future wireless systems. By strategically deploying an IRS and configuring its passive reflection, favorable wireless channel conditions can be created to enhance desired signals, suppress co-channel and inter-cell interference, and refine the channel statistics~\cite{ref_Wu_Tut,ref_Zheng_Sur}. Also, IRS can achieve directional signal suppression by destructively combining the signals reflected by the IRS with those propagated through other paths. The potential of IRS for directional signal suppression has recently been applied for anti-detection applications in~\cite{ref_Zheng_TSP} and~\cite{ref_Xiong_WCL}, where an IRS mounted on a target surface synergizes its reflection pattern with imperfect electromagnetic wave absorbing materials to neutralize residual echo signals in single- and multi-radar systems. Thanks to its ability to dynamically modulate its reflection properties, an IRS offers flexible and real-time ES performance, in stark contrast to traditional ES methods based on materials with fixed properties~\cite{ref_Zheng_Mag}. However, the adaptive ES strategies proposed in~\cite{ref_Zheng_TSP} and~\cite{ref_Xiong_WCL} require accurate knowledge of the positions/directions of all radars, which necessitates embedding additional sensing devices, e.g., rotatable antennas~\cite{ref_RAmodeling}, on the target and thus increases the hardware cost and complexity. In such cases, ES performance can be significantly compromised if the positions of the radars are uncertain or frequently change within a large region.

An effective solution to address the above issue is to establish a shielded zone aligned with the area in which the radars may be located, rendering sensing of the radar positions unnecessary. To establish such a shielded zone, the IRS reflection should be designed to consistently neutralize the radar probing signals echoed back from the target across the entire area where the radars may be located, which we refer to as the unauthorized detection region. Motivated by the above discussion, we propose a new IRS-aided ES strategy to evade potential detection from radars randomly located in an unauthorized detection region. Specifically, an IRS passive reflection optimization problem is formulated to minimize the maximum received signal-to-noise ratio (SNR) of the target echo signal within the region. The problem involves semi-infinite reflection gain constraints, which pose a challenge in obtaining the optimal IRS reflection. By first discretizing the continuous spatial frequency deviation and then leveraging the Lagrange dual method, a semi-closed-form solution for the IRS reflection vector is obtained. Simulation results are provided to demonstrate the ES performance of the proposed IRS-aided strategy compared to other benchmark schemes.

\section{System Model and Problem Formulation}
\begin{figure}[!t]
  \centering
  \includegraphics[width=2.5in]{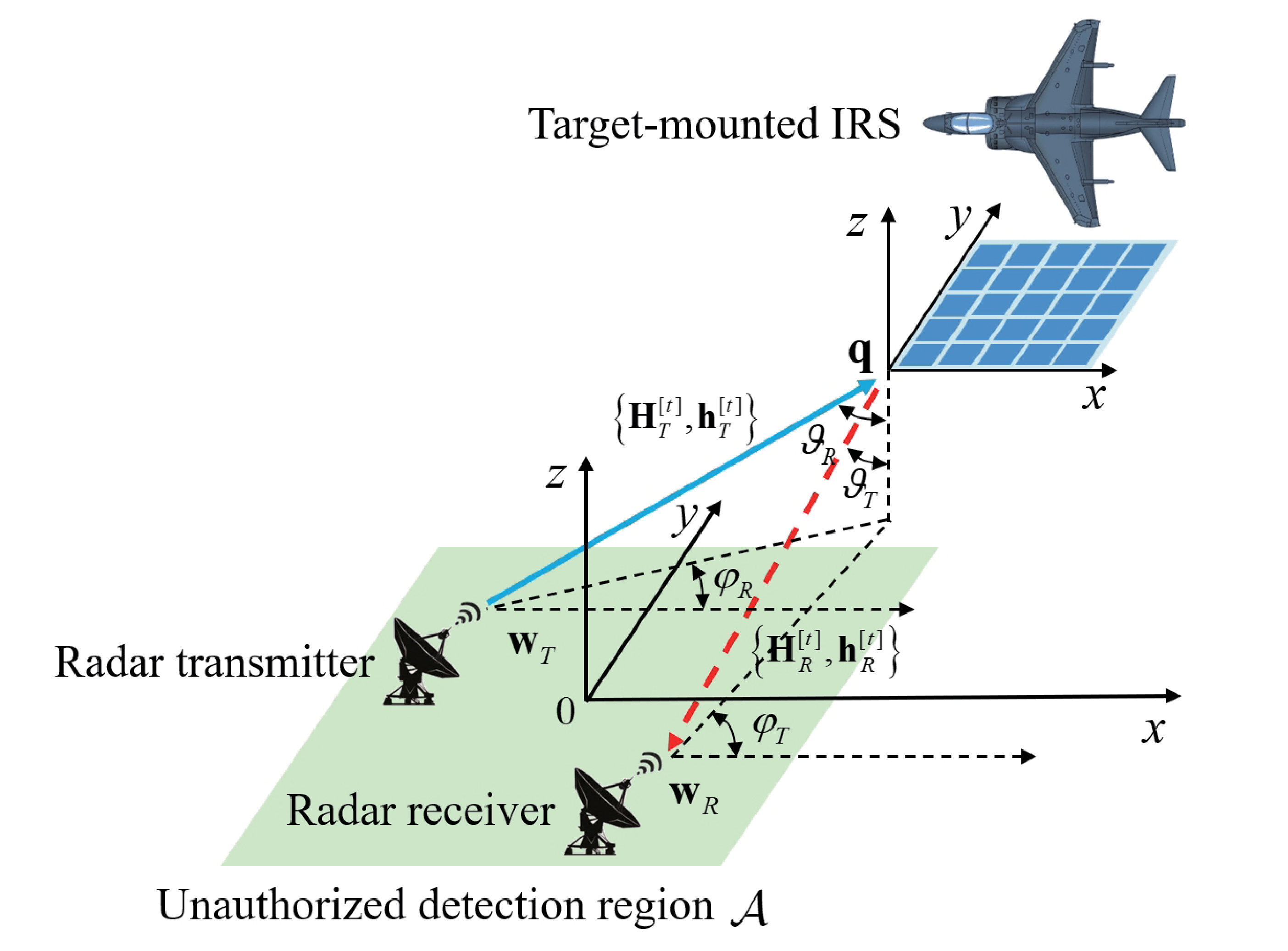} \vspace{-0.2cm}
  \caption{IRS-aided ES system for an unauthorized detection region.}
  \label{fig_system} \vspace{-0.4cm}
\end{figure}

To evade potential detection from radars randomly located in an unauthorized detection region $\mathcal{A}$, we consider an IRS-aided ES system as illustrated in Fig.~\ref{fig_system}, where an IRS is mounted on the aerial target's surface to prevent radar reflections in certain directions. For ease of exposition, we assume that $\mathcal{A}$ is a rectangular region in the $x$-$y$ plane of a three-dimensional (3D) Cartesian coordinate system. We consider a bistatic radar scenario where the radar transmitter and receiver are both equipped with $M$ antennas and located randomly in $\mathcal{A}$ in an attempt to detect the target. The 3D coordinates of the radar transmitter and receiver are denoted by $\mathbf{w}_T \in \mathbb{R}^{3\times 1}$ and $\mathbf{w}_R \in \mathbb{R}^{3\times 1}$, respectively. The target-mounted IRS is parallel to the $x$-$y$ plane and consists of a sub-wavelength uniform planar array (UPA) with $N = N_x N_y$ passive reflecting elements, where $N_x$ and $N_y$ denote the number of reflecting elements along the $x$- and $y$-axes, respectively.\footnote{The proposed IRS-aided ES system can be extended to a non-uniform array setting, where the reflecting elements are non-uniformly deployed to match the irregular target surface. This changes the definition of the IRS steering vector below, but otherwise does not impact implementation of the proposed algorithm.}

Due to the high altitude of the aerial target, the propagation channels between each radar to the IRS can be characterized by a far-field line-of-sight (LoS) model. Furthermore, we focus on the radar detection during one coherent-processing interval (CPI) $T_c$, during which the channels and geometry-related parameters are assumed to be constant. The position of the target/IRS during the CPI is denoted by $\mathbf{q} \in \mathbb{R}^{3\times 1}$. 

We define the one-dimensional (1D) steering vector for a uniform linear array (ULA) as
\begin{equation}
  \label{deqn_ex1a}
   \mathbf{e}(\phi,\bar{N}) \triangleq \left [1, e^{-j\pi \phi}, \dots, e^{-j\pi(\bar{N}-1)\phi} \right ]^T,
\end{equation}
where $\phi \in \left [ 0,2\pi \right )$ denotes the constant phase-shift difference between the signals at two adjacent antennas/elements and $\bar{N}$ denotes the number of antennas/elements in the ULA. Denoting $\vartheta_R(\mathbf{q},\mathbf{w}_T)$ and $\varphi_R(\mathbf{q},\mathbf{w}_T)$ as respectively the zenith and azimuth angles-of-arrival (AoAs) for the transmit link from the radar transmitter $\mathbf{w}_T \in \mathcal{A}$ to the IRS, the receive array response vector of the IRS can be expressed as
\begin{align}
  \mathbf{a}_R(\mathbf{q},\mathbf{w}_T) = & \mathbf{e}\left (d_e \bar{\Phi}_R(\mathbf{q},\mathbf{w}_T),N_x \right ) \nonumber\\
  & \otimes \mathbf{e} \left (d_e \bar{\Omega}_R(\mathbf{q},\mathbf{w}_T),N_y \right ) \in \mathbb{C}^{N \times 1}, \label{deqn_ex2a}
\end{align}
where $d_e \triangleq \frac{2\Delta_e}{\lambda}$, $\lambda$ is the probing signal wavelength, $\Delta_e \leq \frac{\lambda}{2}$ denotes the element spacing at the IRS, $\bar{\Phi}_R(\mathbf{q},\mathbf{w}_T) \triangleq \mathrm{sin}(\vartheta_R(\mathbf{q},\mathbf{w}_T)) \mathrm{cos}(\varphi_R(\mathbf{q},\mathbf{w}_T))$ and $\bar{\Omega}_R(\mathbf{q},\mathbf{w}_T) \triangleq \mathrm{sin}(\vartheta_R(\mathbf{q},\mathbf{w}_T)) \mathrm{sin}(\varphi_R(\mathbf{q},\mathbf{w}_T))$ are the \textit{spatial frequencies} along the $x$- and $y$-dimensions corresponding to the AoAs, respectively, and $\otimes$ denotes the Kronecker product. The array response of the radar transmitter $\bar{\mathbf{a}}_{T}(\mathbf{q},\mathbf{w}_T) \in \mathbb{C}^{M \times 1}$ can be similarly obtained.

Due to the movement of the target, the propagation links are subject to a Doppler frequency shift. The $N\times M$ channel from the transmit radar to the IRS at time $t$ with $0\leq t\leq T_c$ is given by
\begin{equation}
  \label{deqn_ex3a}
  \hspace{-0.1cm}\mathbf{H}_T^{[t]}(\mathbf{q},\mathbf{w}_T) = \rho_T(\mathbf{q},\mathbf{w}_T) e^{j2\pi f_T t} \mathbf{a}_R(\mathbf{q},\mathbf{w}_T) \bar{\mathbf{a}}_{T}^H(\mathbf{q},\mathbf{w}_T),
\end{equation}
where $\rho_T(\mathbf{q},\mathbf{w}_T) \triangleq \frac{\sqrt{\alpha}}{d_T(\mathbf{q},\mathbf{w}_T)} e^{-j\frac{2\pi}{\lambda}d_T(\mathbf{q},\mathbf{w}_T)}$ is the complex-valued path gain, $d_T(\mathbf{q},\mathbf{w}_T) = \| \mathbf{q} - \mathbf{w}_T \|$ is the propagation distance between the radar transmitter and IRS, $\alpha$ is the path gain at a reference distance of 1 meter (m), $f_T \triangleq v\mathrm{cos}(\vartheta_R(\mathbf{q},\mathbf{w}_T)) \mathrm{cos}(\varphi_R(\mathbf{q},\mathbf{w}_T))/\lambda$ denotes the Doppler frequency of the transmit link, and $v$ is the speed of the aerial target. The far-field LoS channel from the radar transmitter to the target at time $t$, denoted as $(\mathbf{h}_T^{[t]}(\mathbf{q},\mathbf{w}_T))^H \in \mathbb{C}^{1 \times M}$, can be expressed as
\begin{equation}
  \label{deqn_ex4a}
  (\mathbf{h}_T^{[t]}(\mathbf{q},\mathbf{w}_T))^H = \rho_T(\mathbf{q},\mathbf{w}_T)  e^{j2\pi f_T t} \bar{\mathbf{a}}_{T}^H(\mathbf{q},\mathbf{w}_T).
\end{equation}

Similarly, denoting $\vartheta_T(\mathbf{q},\mathbf{w}_R)$ and $\varphi_T(\mathbf{q},\mathbf{w}_R)$ as respectively the zenith and azimuth angles-of-departure (AoDs) for the receive link from the IRS to the radar receiver $\mathbf{w}_R \in \mathcal{A}$, the reflect array response vector at the IRS can be obtained as
\begin{align}
  \mathbf{a}_T(\mathbf{q},\mathbf{w}_R) = & \mathbf{e}\left (d_e \bar{\Phi}_T(\mathbf{q},\mathbf{w}_R),N_x \right )\nonumber\\
  &  \otimes \mathbf{e} \left (d_e \bar{\Omega}_T(\mathbf{q},\mathbf{w}_R),N_y \right ) \in \mathbb{C}^{N \times 1},\label{deqn_ex5a}
\end{align}
where $\bar{\Phi}_T(\mathbf{q},\mathbf{w}_R) \triangleq \mathrm{sin}(\vartheta_T(\mathbf{q},\mathbf{w}_R)) \mathrm{cos}(\varphi_T(\mathbf{q},\mathbf{w}_R))$ and $\bar{\Omega}_T(\mathbf{q},\mathbf{w}_R) \triangleq \mathrm{sin}(\vartheta_T(\mathbf{q},\mathbf{w}_R)) \mathrm{sin}(\varphi_T(\mathbf{q},\mathbf{w}_R))$ denote the spatial frequencies along the $x$- and $y$-dimensions corresponding to the AoDs, respectively. Then, the channels from the IRS and target to the radar receiver at time $t$, denoted as $\mathbf{H}_R^{[t]}(\mathbf{q},\mathbf{w}_R) \in \mathbb{C}^{M \times N}$ and $\mathbf{h}_R^{[t]}(\mathbf{q},\mathbf{w}_R) \in \mathbb{C}^{M \times 1}$, can be expressed similarly to \eqref{deqn_ex3a} and \eqref{deqn_ex4a}, i.e.,
\begin{align}
  \hspace{-0.15cm}\mathbf{H}_R^{[t]}(\mathbf{q},\mathbf{w}_R)&=\rho_R(\mathbf{q},\mathbf{w}_R)  e^{j2\pi f_R t} \bar{\mathbf{a}}_{R}(\mathbf{q},\mathbf{w}_R) \mathbf{a}_{T}^H(\mathbf{q},\mathbf{w}_R), \label{deqn_ex6a}\\
  \hspace{-0.15cm}\mathbf{h}_R^{[t]}(\mathbf{q},\mathbf{w}_R)&=\rho_R(\mathbf{q},\mathbf{w}_R)  e^{j2\pi f_R t} \bar{\mathbf{a}}_{R}(\mathbf{q},\mathbf{w}_R), \label{deqn_ex7a}
\end{align}
where $\rho_R(\mathbf{q},\mathbf{w}_R) \triangleq \frac{\sqrt{\alpha}}{d_R(\mathbf{q},\mathbf{w}_R)} e^{-j\frac{2\pi}{\lambda}d_R(\mathbf{q},\mathbf{w}_R)}$ is the complex-valued path gain, $d_R(\mathbf{q},\mathbf{w}_R) = \| \mathbf{q} - \mathbf{w}_R \|$ is the propagation distance between the IRS and radar receiver, $f_R \triangleq v\mathrm{cos}(\vartheta_T(\mathbf{q},\mathbf{w}_R)) \mathrm{cos}(\varphi_T(\mathbf{q},\mathbf{w}_R))/\lambda$ denotes the Doppler frequency of the receive link, and $\bar{\mathbf{a}}_{R}(\mathbf{q},\mathbf{w}_R) \in \mathbb{C}^{M \times 1}$ is the array response of the radar receiver.

Denoting the pulsed waveform vector of the radar transmitter as $\mathbf{x}^{[t]} \in \mathbb{C}^{M \times 1}$, the echoes from the target/IRS at the radar receiver at time $t$ can be characterized as
\begin{align}
  \mathbf{y}^{[t]}&(\mathbf{q},\mathbf{w}_T,\mathbf{w}_R,\bm{\theta}^{[t]}) = \underbrace{\mathbf{H}_R^{[t]}(\mathbf{q},\mathbf{w}_R) \mathrm{diag}(\bm{\theta}^{[t]}) \mathbf{H}_T^{[t]}(\mathbf{q},\mathbf{w}_T) \mathbf{x}^{[t]}}_{\mathrm{Reflected \; signal \; from \; IRS}} \nonumber\\
  & \quad \quad \quad \quad + \underbrace{\tau_S \mathbf{h}_R^{[t]}(\mathbf{q},\mathbf{w}_R) (\mathbf{h}_T^{[t]}(\mathbf{q},\mathbf{w}_T))^H \mathbf{x}^{[t]}}_{\mathrm{Reflected \; signal \; from \; target}} + \mathbf{n}^{[t]}, \label{deqn_ex8a}
\end{align}
where $\bm{\theta}^{[t]} \triangleq \left[\theta_1^{[t]}, \theta_2^{[t]}, \dots, \theta_N^{[t]}\right]^T$ denotes the IRS reflection vector, $\theta_n^{[t]}=\beta_n^{[t]} e^{j\psi_n^{[t]}}, n=(n_x-1)N_y+n_y$ corresponds to the IRS element at position $(n_x, n_y)$, $\beta_n^{[t]} \in \left[0,1\right]$ and $\phi_n^{[t]} \in \left [0,2\pi\right )$ are respectively the reflection amplitude and phase shift of the $n$-th element at time $t$, $\tau_S$ denotes the isotropic complex-valued RCS of the target surface, and $\mathbf{n}^{[t]} \sim \mathcal{N}_c(\mathbf{0},\sigma^2 \mathbf{I}_M)$ is zero-mean additive white Gaussian noise (AWGN) with variance of $\sigma^2$.

Based on \eqref{deqn_ex8a}, the SNR at the radar receiver at time $t$ is given by
\vspace{-0.1cm}
\begin{align}
  \gamma^{[t]}(\mathbf{q},\mathbf{w}_T,\mathbf{w}_R,\bm{\theta}^{[t]}) = & \frac{1}{M\sigma^2} \|\rho_R(\mathbf{q},\mathbf{w}_R)  e^{j2\pi f_R t} \bar{\mathbf{a}}_{R}(\mathbf{q},\mathbf{w}_R) \times \nonumber\\
  R^{[t]}(\mathbf{q},\mathbf{w}_T,\mathbf{w}_R,\bm{\theta}^{[t]})& \bar{\mathbf{a}}_{T}^H(\mathbf{q},\mathbf{w}_T) \mathbf{x}^{[t]} \rho_T(\mathbf{q},\mathbf{w}_T)  e^{j2\pi f_T t} \|^2 \nonumber\\
  = & \frac{G^{[t]}}{M\sigma^2} \|R^{[t]}(\mathbf{q},\mathbf{w}_T,\mathbf{w}_R,\bm{\theta}^{[t]})\|^2, \label{deqn_ex9a}
\end{align}
where $R^{[t]}(\mathbf{q},\mathbf{w}_T,\mathbf{w}_R,\bm{\theta}^{[t]}) \triangleq \mathbf{a}_{T}^H(\mathbf{q},\mathbf{w}_R) \mathrm{diag}(\bm{\theta}^{[t]}) \times \mathbf{a}_R(\mathbf{q},\mathbf{w}_T) + \tau_S$ denotes the complex-valued reflection gain at the target/IRS that depends on the AoAs and AoDs of the target/IRS and the IRS reflection and $G^{[t]} = M \| \rho_R(\mathbf{q},\mathbf{w}_R) \rho_T(\mathbf{q},\mathbf{w}_T) \bar{\mathbf{a}}_{T}^H(\mathbf{q},\mathbf{w}_T) \mathbf{x}^{[t]}\|^2$ is the normalized receive signal power when $\|R^{[t]}(\mathbf{q},\mathbf{w}_T,\mathbf{w}_R,\bm{\theta}^{[t]})\|^2 = 1$.

To reduce the radar detection probability, our objective is to minimize the maximum SNR within the unauthorized detection region $\mathcal{A}$ by optimizing the IRS reflection vector $\bm{\theta}^{[t]}$ for a given target position $\mathbf{q}$. The problem is formulated as\footnote{To achieve real-time IRS reflection adjustment, the geometric information between the target and the unauthorized detection region can be predicted and continuously updated based on prior knowledge of the target's trajectory. Additionally, the optimized $\bm{\theta}^{[t]}$ for all possible radar positions can be precalculated offline and stored in a database on the IRS controller.}
\vspace{-0.2cm}
\begin{subequations}\label{eq:1}
  \begin{alignat}{2}
    \mathrm{(P1)} \quad \min_{\bm{\theta}^{[t]}} \quad & \max_{\mathbf{w}_T, \mathbf{w}_R \in \mathcal{A}} \| R^{[t]}(\mathbf{q},\mathbf{w}_T,\mathbf{w}_R,\bm{\theta}^{[t]}) \|^2 & \label{eq:1A}\\
    \mbox{s.t.} \quad 
    & | \theta_n^{[t]} | \leq 1, \forall n = 1, 2, \dots, N. \label{eq:1B}
  \end{alignat}
\end{subequations}
Note that $\frac{G^{[t]}}{M\sigma^2}$ in \eqref{deqn_ex9a} is a constant independent of $\bm{\theta}^{[t]}$ and thus can be omitted in the objective function.

\section{IRS Passive Reflection Optimization}
According to \eqref{deqn_ex2a} and \eqref{deqn_ex5a}, the reflection gain defined in \eqref{deqn_ex9a} can be rewritten as\footnote{Since $R^{[t]}(\mathbf{q},\mathbf{w}_T,\mathbf{w}_R,\bm{\theta}^{[t]})$ is time varying only because of ${\theta}^{[t]}$, we omit the time index $[t]$ for brevity in the following.}
\begin{align}
  & \| R(\mathbf{q},\mathbf{w}_T,\mathbf{w}_R,\bm{\theta}) \|^2 \nonumber\\
  = & \| \tau_S + \left[\mathbf{a}_{T}^H(\mathbf{q},\mathbf{w}_R) \odot \mathbf{a}_R^{T}(\mathbf{q},\mathbf{w}_T)\right] \bm{\theta} \|^2 \nonumber\\
  = & \| \tau_S + \left[ \mathbf{e}^T\left (d_e (\bar{\Phi}_R(\mathbf{q},\mathbf{w}_T) - \bar{\Phi}_T(\mathbf{q},\mathbf{w}_R)),N_x \right ) \right. \nonumber\\
  & \left. \quad \otimes \mathbf{e}^T\left (d_e (\bar{\Omega}_R(\mathbf{q},\mathbf{w}_T) - \bar{\Omega}_T(\mathbf{q},\mathbf{w}_R)),N_y \right ) \right] \bm{\theta} \|^2, \label{deqn_ex1b}
\end{align}
where $\odot$ denotes the Hadamard product. For any given target position $\mathbf{q}$, we define $\Phi_{\mathrm{min}}$ and $\Phi_{\mathrm{max}}$ as respectively the minimum and maximum difference between the spatial frequency $\bar{\Phi}_R(\mathbf{q},\mathbf{w}_T)$ associated with the AoAs and $\bar{\Phi}_T(\mathbf{q},\mathbf{w}_R)$ associated with the AoDs along the $x$-axis in unauthorized detection region $\mathcal{A}$, i.e.,
\begin{subequations}\label{deqn_ex2b}
  \begin{align}
    \Phi_{\mathrm{min}} & \triangleq \min_{\mathbf{w}_T, \mathbf{w}_R \in \mathcal{A}} \bar{\Phi}_R(\mathbf{q},\mathbf{w}_T) - \bar{\Phi}_T(\mathbf{q},\mathbf{w}_R), \label{deqn_ex2b_A}\\
    \Phi_{\mathrm{max}} & \triangleq \max_{\mathbf{w}_T, \mathbf{w}_R \in \mathcal{A}} \bar{\Phi}_R(\mathbf{q},\mathbf{w}_T) - \bar{\Phi}_T(\mathbf{q},\mathbf{w}_R). \label{deqn_ex2b_B}
  \end{align}
\end{subequations}
The minimum and maximum deviation of spatial frequency $\bar{\Omega}_R(\mathbf{q},\mathbf{w}_T)$ from $\bar{\Omega}_T(\mathbf{q},\mathbf{w}_R)$ along the $y$-axis, denoted as $\Omega_{\mathrm{min}}$ and $\Omega_{\mathrm{max}}$, respectively, can be similarly obtained. If we define $\Phi \triangleq \bar{\Phi}_R(\mathbf{q},\mathbf{w}_T) - \bar{\Phi}_T(\mathbf{q},\mathbf{w}_R)$ and $\Omega \triangleq \bar{\Omega}_R(\mathbf{q},\mathbf{w}_T) - \bar{\Omega}_T(\mathbf{q},\mathbf{w}_R)$, the unauthorized detection region $\mathcal{A}$ can be equivalently described by $\mathbb{A}\triangleq \{(\Phi,\Omega) | \Phi \in \left[\Phi_{\mathrm{min}}, \Phi_{\mathrm{max}} \right], \Omega \in \left[\Omega_{\mathrm{min}}, \Omega_{\mathrm{max}} \right] \}$ in the angular domain. By introducing the slack optimization variable $\eta$ to denote the maximum reflection gain within $\mathcal{A}$, (P1) can be equivalently written as
\vspace{-0.1cm}
\begin{subequations}\label{eq:2}
  \begin{alignat}{2}
    \mathrm{(P2)} \quad \min_{\eta,\bm{\theta}} \quad & \eta & \label{eq:2A}\\
    \mbox{s.t.} \quad 
    & \left|\left| \left[\mathbf{e}^T\left(d_e \Phi,N_x \right) \otimes \mathbf{e}^T\left(d_e \Omega,N_y \right)\right] \bm{\theta} + \tau_S \right|\right|^2 \nonumber\\
    & \quad \leq \eta, \; (\Phi,\Omega) \in \mathbb{A}, \label{eq:2B}\\
    & \theta_n^{\ast}\theta_n \leq 1, \forall n = 1, 2, \dots, N. \label{eq:2C}
  \end{alignat}
\end{subequations}
For the above problem, $\bm{\theta}$ should be designed such that the reflection gain $\| \left[\mathbf{e}^T\left(d_e \Phi,N_x \right) \otimes \mathbf{e}^T\left(d_e \Omega,N_y \right)\right] \bm{\theta} + \tau_S \|^2 $ is approximately equivalent for all spatial frequency pairs $(\Phi,\Omega)$ within angular domain $\mathbb{A}$. The reflection gain constraint \eqref{eq:2B} involves semi-infinite constraints that render a direct solution to problem (P2) intractable.

To address this challenge, we first discretize the continuous spatial frequency deviation by sampling $K$ points within angular domain $\mathbb{A}$. In particular, $\mathbb{A}$ is discretized as $ \{(\Phi_k,\Omega_k)\}_{k=1}^{K}$, where $\Phi_k \in \left[\Phi_{\mathrm{min}}, \Phi_{\mathrm{max}} \right]$ and $\Omega_k \in \left[\Omega_{\mathrm{min}}, \Omega_{\mathrm{max}} \right]$ with $k = 1,2,\dots,K$ respectively denote the spatial frequency deviations along the $x$- and $y$-dimensions corresponding to the $k$-th sampling point. Problem (P2) can thus be approximated as
\vspace{-0.2cm}
\begin{subequations}\label{eq:3}
  \begin{alignat}{2}
    \mathrm{(P3)} \quad \min_{\eta,\bm{\theta}} \quad & \eta & \label{eq:3A}\\
    \mbox{s.t.} \quad 
    & \| \mathbf{u}_k^T \bm{\theta} + \tau_S \|^2 \leq \eta, \; \forall k = 1,2,\dots,K, \label{eq:3B}\\
    & \eqref{eq:2C}, \label{eq:3C}
  \end{alignat}
\end{subequations}
where $\mathbf{u}_k \triangleq \mathbf{e} (d_e \Phi_k,N_x ) \otimes \mathbf{e} (d_e \Omega_k,N_y )$ is the array response vector corresponding to the $k$-th sampling point within angular domain $\mathbb{A}$. Note that (P3) is equivalent to (P2) if $K \to \infty$ and the sampling points cover the entire angular domain $\mathbb{A}$. Since (P3) is convex and satisfies Slater's condition, \textit{strong duality} holds between (P3) and its Lagrange dual problem. Therefore, we can optimally solve (P3) by exploiting the Lagrange dual method.

Let $\bm{\lambda} \triangleq \{\lambda_1,\lambda_2,\dots,\lambda_K\}$ and $\bm{\mu} \triangleq \{\mu_1,\mu_2,\dots,\mu_{N}\}$ denote the non-negative dual variables associated with constraints \eqref{eq:3B} and \eqref{eq:2C}, respectively. Then, the Lagrangian function associated with (P3) is given as
\vspace{-0.2cm}
\begin{align}
  \mathcal{L}(\eta,\bm{\theta},\bm{\lambda},\bm{\mu}) = & \bm{\theta}^H \mathbf{Q} \bm{\theta} + {\sum_{k=1}^{K}{\lambda_k \left(\tau_S \bm{\theta}^H \mathbf{u}_k + \tau_S^{\ast} \mathbf{u}_k^{H} \bm{\theta} + |\tau_S|^2 \right)}} \nonumber\\
  & + \eta \left(1 - {\sum_{k=1}^{K}{\lambda_k}} \right) - {\sum_{n=1}^{N}{\mu_n}}, \label{deqn_ex3b}
\end{align}
where $\mathbf{Q} \triangleq \sum_{k=1}^{K}{\lambda_k \mathbf{u}_k \mathbf{u}_k^H + \mathrm{diag}(\bm{\mu})}$. Accordingly, the dual function of (P3) is given by
\begin{equation}
  \label{deqn_ex4b}
  f(\bm{\lambda},\bm{\mu}) = \min_{\eta,\bm{\theta}} \mathcal{L}(\eta,\bm{\theta},\bm{\lambda},\bm{\mu}).
\end{equation}
For the dual function $f(\bm{\lambda},\bm{\mu})$ to be lower-bounded (i.e., $f(\bm{\lambda},\bm{\mu}) > -\infty$), $\sum_{k=1}^{K}{\lambda_k} = 1$ must hold. Otherwise, setting $\eta \to \infty$ (or $\eta \to -\infty$) results in $f(\bm{\lambda},\bm{\mu}) \to -\infty$ if $\sum_{k=1}^{K}{\lambda_k} > 1$ (or $\sum_{k=1}^{K}{\lambda_k} < 1$). Thus, the dual problem for (P3) is given by
\begin{subequations}\label{eq:4}
  \begin{alignat}{2}
    \mathrm{(D3)} \quad \max_{\bm{\lambda},\bm{\mu}} \quad & f(\bm{\lambda},\bm{\mu}) & \label{eq:4A}\\
    \mbox{s.t.} \quad 
    & {\textstyle \sum_{k=1}^{K}{\lambda_k}} = 1, \label{eq:4B}\\
    & \lambda_k \geq 0,\;k = 1,2,\dots,K, \label{eq:4C}\\
    & \mu_n \geq 0,\;n = 1,2,\dots,N. \label{eq:4D}
  \end{alignat}
\end{subequations}

In the following, we first solve problem \eqref{deqn_ex4b} to obtain $f(\bm{\lambda},\bm{\mu})$ for any given feasible dual variables $\{\bm{\lambda},\bm{\mu}\}$, then solve (D3) to obtain the optimal $\{\bm{\lambda},\bm{\mu}\}$ to maximize $f(\bm{\lambda},\bm{\mu})$, and finally construct the optimal primal solution to (P3).

\textit{1) Obtaining $f(\bm{\lambda},\bm{\mu})$ by Solving Problem \eqref{deqn_ex4b} for Given $\{\bm{\lambda},\bm{\mu}\}$:} For any given $\{\bm{\lambda},\bm{\mu}\}$, problem \eqref{deqn_ex4b} can be decomposed into the following two subproblems.
\begin{align}
  & \min_{\eta}\eta\left(1 - {\textstyle \sum_{k=1}^{K}{\lambda_k}} \right), \label{deqn_ex5b}\\
  & \min_{\bm{\theta}} \bm{\theta}^H \mathbf{Q} \bm{\theta} + {\textstyle \sum_{k=1}^{K}{\lambda_k \left(\tau_S \bm{\theta}^H \mathbf{u}_k + \tau_S^{\ast} \mathbf{u}_k^{H} \bm{\theta} \right)}}. \label{deqn_ex6b}
\end{align}
Denote by $\eta^{(\lambda,\mu)}$ and $\bm{\theta}^{(\lambda,\mu)}$ the optimal solutions to \eqref{deqn_ex5b} and \eqref{deqn_ex6b}, respectively. For problem \eqref{deqn_ex5b}, since $1 - \sum_{k=1}^{K}{\lambda_k} = 0$ holds for any given feasible dual variables, the value of the objective is always zero. Thus, we can choose any arbitrary real number as the optimal solution $\eta^{(\lambda,\mu)}$. For problem \eqref{deqn_ex6b}, we set the first order derivative of the objective function with respect to $\bm{\theta}$ to zero, i.e., $\mathbf{Q} \bm{\theta} + \tau_S \sum_{k=1}^{K}{\lambda_k \mathbf{u}_k} = \mathbf{0}$, and the optimal solution to problem \eqref{deqn_ex6b} can thus be obtained as
\begin{equation}
  \label{deqn_ex7b}
  \bm{\theta}^{(\lambda,\mu)} = -\tau_S {\textstyle\sum_{k=1}^{K}{\lambda_k \mathbf{Q}^{-1} \mathbf{u}_k}}.
\end{equation}

\textit{2) Finding Optimal Dual Solution to (D3):} With $\eta^{(\lambda,\mu)}$ and $\bm{\theta}^{(\lambda,\mu)}$ obtained, we then solve the dual problem (D3) to find the optimal $\{\bm{\lambda},\bm{\mu}\}$ to maximize $f(\bm{\lambda},\bm{\mu})$. According to $\sum_{k=1}^{K}{\lambda_k^{\star}} = 1$ and by substituting $\eta^{(\lambda,\mu)}$ and $\bm{\theta}^{(\lambda,\mu)}$ into $f(\bm{\lambda},\bm{\mu})$, we have
\begin{equation}
  \label{deqn_ex8b}
  f(\bm{\lambda},\bm{\mu}) = -|\tau_S|^2 \mathbf{v}^H \mathbf{Q}^{-1} \mathbf{v} - \mathbf{1}_{N_x}^T \bm{\mu} + |\tau_S|^2,
\end{equation}
where $\mathbf{v} \triangleq \sum_{k=1}^{K}{\lambda_k \mathbf{u}_k}$. Furthermore, by applying the Schur complement, the dual problem can be transformed into an equivalent semidefinite optimization problem as follows:
\begin{subequations}\label{eq:4}
  \begin{alignat}{2}
    \mathrm{(P4)} \quad \max_{q,\bm{\lambda},\bm{\mu}} \quad & q & \label{eq:5A}\\
    \mbox{s.t.} \quad 
    & \begin{bmatrix}
      |\tau_S|^2 - \mathbf{1}_{N_x}^T \bm{\mu} - q & \tau_S^{\ast} \mathbf{v}^H\\
      \tau_S \mathbf{v} & \mathbf{Q}
    \end{bmatrix} \succeq \mathbf{0}, \label{eq:5B}\\
    & \eqref{eq:4B} - \eqref{eq:4D}. \label{eq:5C}
  \end{alignat}
\end{subequations}
Problem (P4) can be effectively solved via semidefinite programming (SDP) or linear matrix inequality (LMI) optimization, with a complexity of order $\mathcal{O}((N+K)^{4.5} \mathrm{ln}(1/\epsilon))$ for a given solution accuracy $\epsilon > 0$~\cite{ref_Luo_SDR}. More sampling points will achieve some performance improvement, but will inevitably increase the complexity, leading to a non-trivial trade-off between ES performance and computational complexity.

\textit{3) Constructing Optimal Primal Solution to (P3):} With optimal dual variables $\bm{\lambda}^{\star}$ and $\bm{\mu}^{\star}$ obtained by solving (D3), the optimal solutions to (P3), denoted as $\bm{\theta}^{\star}$ and $\eta^{\star}$, can be expressed as\footnote{Based on our extensive simulation experiments, most of the optimized amplitudes of the reflecting elements tend to be equal to one, i.e., $|\theta_n^{\star}| = 1$. For $|\theta_n^{\star}| < 1$, the reflecting element absorbs part of the received electromagnetic wave energy to reduce the effective RCS of the target~\cite{ref_IRShybrid}.}
\begin{align}
  \bm{\theta}^{\star} & = -\tau_S {\sum_{k=1}^{K}{\lambda_k^{\star} \left({\sum_{i=1}^{K}{\lambda_i^{\star} \mathbf{u}_i \mathbf{u}_i^H + \mathrm{diag}(\bm{\mu}^{\star})}} \right)^{-1} \mathbf{u}_k}}, \label{deqn_ex9b}\\
  \eta^{\star} & = \max_{k=1,2,\dots,K} \| \mathbf{u}_k^T \bm{\theta}^{\star} + \tau_S \|^2. \label{deqn_ex10b}
\end{align}

\textit{Remark:} Based on the Karush-Kuhn-Tucker (KKT) conditions~\cite{ref_Boyd}, the complementary slackness condition corresponding to constraint \eqref{eq:3B} can be expressed as
\begin{equation}
  \label{deqn_ex11b}
  \lambda_k^{\star} \left( \| \mathbf{u}_k^T \bm{\theta}^{\star} + \tau_S \|^2 - \eta^{\star} \right) = 0,\; \forall k = 1,2,\dots,K.
\end{equation}
For any given sampling index $k$, if $\lambda_k^{\star} > 0$, $\| \mathbf{u}_k^T \bm{\theta}^{\star} + \tau_S \|^2 = \eta^{\star}$ must hold. Conversely, if $\| \mathbf{u}_k^T \bm{\theta}^{\star} + \tau_S \|^2 < \eta^{\star}$, to satisfy complementary slackness condition \eqref{deqn_ex11b}, $\lambda_k^{\star} = 0$ must be true. As we can observe from \eqref{deqn_ex9b}, if $\lambda_k^{\star} = 0$, the array response vector $\mathbf{u}_k$ corresponding to the $k$-th sampling point within angular domain $\mathbb{A}$ has no impact on the optimal reflection vector $\bm{\theta}^{\star}$. If we define $\mathcal{K} \triangleq \{k | \lambda_k^{\star}>0, k=1,2,\dots,K \}$, there are $|\mathcal{K}|$ effective sampling points for optimizing the IRS reflection. This phenomenon indicates that in addition to the number of sampling points, their distribution also affects the ES performance and computational complexity.

\section{Simulation Results}
In this section, we present simulation results to demonstrate the performance of the proposed IRS-aided ES system. To facilitate a more intuitive presentation of the variation in reflection gain across the angular domain, we consider a ULA-based IRS, i.e., $N_y = 1$ and $N = N_x$. Thus, only the horizontal reflection gain for spatial frequency $\Phi$ along the $x$-dimension is considered, and angle domain $\mathbb{A}$ reduces to a spatial frequnecy deviation interval $\left[\Phi_{\mathrm{min}}, \Phi_{\mathrm{max}} \right]$. Unless otherwise stated, we assume that the radar system operates at 2 GHz with wavelength $\lambda =$ 0.15 m; the separation between adjacent elements at the IRS is set as $\Delta_e = \frac{\lambda}{2} =$ 0.075 m; the minimum and maximum deviation of the spatial frequencies along the $x$-axis in the unauthorized detection region are set as $\Phi_{\mathrm{min}} = -$0.25 and $\Phi_{\mathrm{max}} =$ 0.25, respectively. According to~\cite{ref_Neunteufel_RCS}, the isotropic complex-valued RCS of the target surface is given by $\tau_S = \frac{4\pi S^2}{\lambda^2}e^{j\xi}$, where the effective echo surface area of the target is set as $S =$ 0.1 $\mathrm{m}^2$ and the phase shift $\xi$ is uniformly distributed in $\left [ 0,2\pi \right )$. The angular region $\left[\Phi_{\mathrm{min}}, \Phi_{\mathrm{max}} \right]$ is discretized with $K = 20$ uniformly spaced sampling points, and the corresponding array response vectors can be expressed as $\mathbf{u}_k = \mathbf{e}(d_e (\Phi_{\mathrm{min}}+(k-1)\bigtriangleup),N_x), k=1,2,\dots,K$ with $\bigtriangleup \triangleq \frac{\Phi_{\mathrm{max}}-\Phi_{\mathrm{min}}}{K-1}$. To validate the effectiveness of our proposed ES scheme, which we refer to as the \textit{Lagrange dual-based optimization}, the following three benchmark schemes are considered:
\begin{itemize}
  \item{\textit{Baseline System without IRS:} The target is detected by the radars without the aid of the IRS. This method can be implemented by setting $\bm{\theta} = \mathbf{0}$ in our proposed algorithm.}
  \item{\textit{Baseline System with a Single Sampling Point:} The IRS reflection vector is designed exploiting the reverse alignment-based solution proposed in \cite{ref_Zheng_TSP} to mitigate the reflection gain for $\Phi = 0$.}
  \item{\textit{Random Phase Shift Design:} The phase shifts of the IRS elements are randomly generated following a uniform distribution on $\left [ 0,2\pi \right )$.}
\end{itemize}


\begin{figure*}
	\setlength{\abovecaptionskip}{-5pt}
	\setlength{\belowcaptionskip}{-10pt}
	\centering
	\begin{minipage}[t]{0.32\linewidth}
		\centering
		\includegraphics[width=2in]{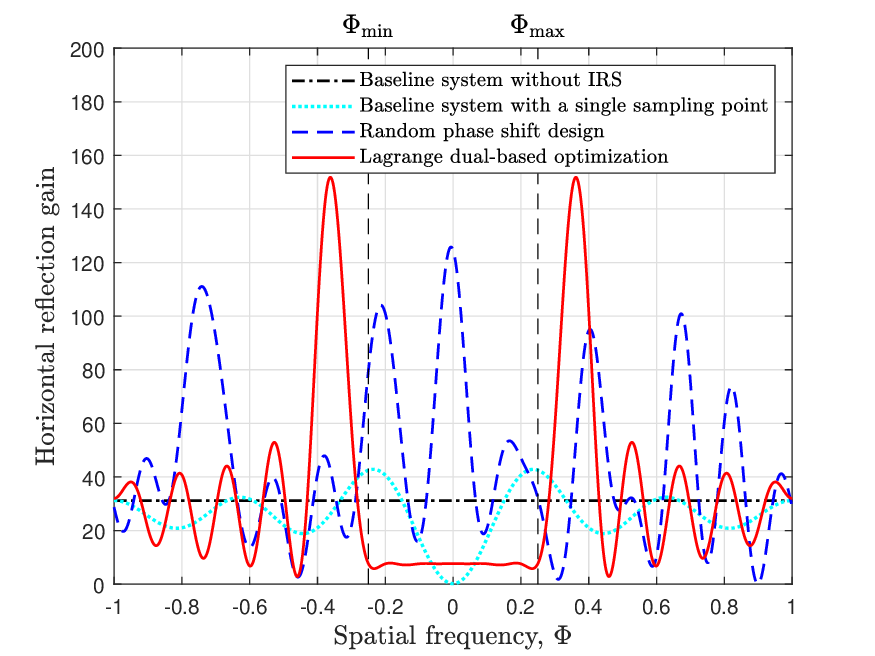}
		\caption{Comparison of the horizontal reflection gain obtained by different methods versus spatial frequency $\Phi$ for $N_x$ = 16.}
		\label{fig_beam}
	\end{minipage}%
	\hspace{0.3cm}\begin{minipage}[t]{0.32\linewidth}
		\centering
		\includegraphics[width=2in]{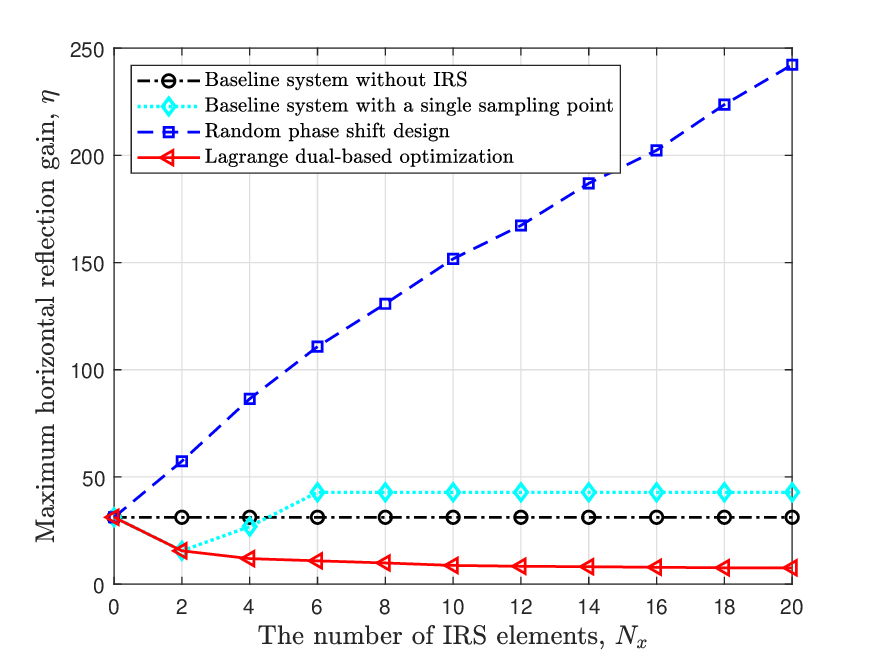}
		\caption{Maximum horizontal reflection gain obtained by different methods versus the number of IRS elements $N_x$.}
		\label{fig_gain}
	\end{minipage}
	\hspace{0.3cm}\begin{minipage}[t]{0.32\linewidth}
		\centering
		\includegraphics[width=2in]{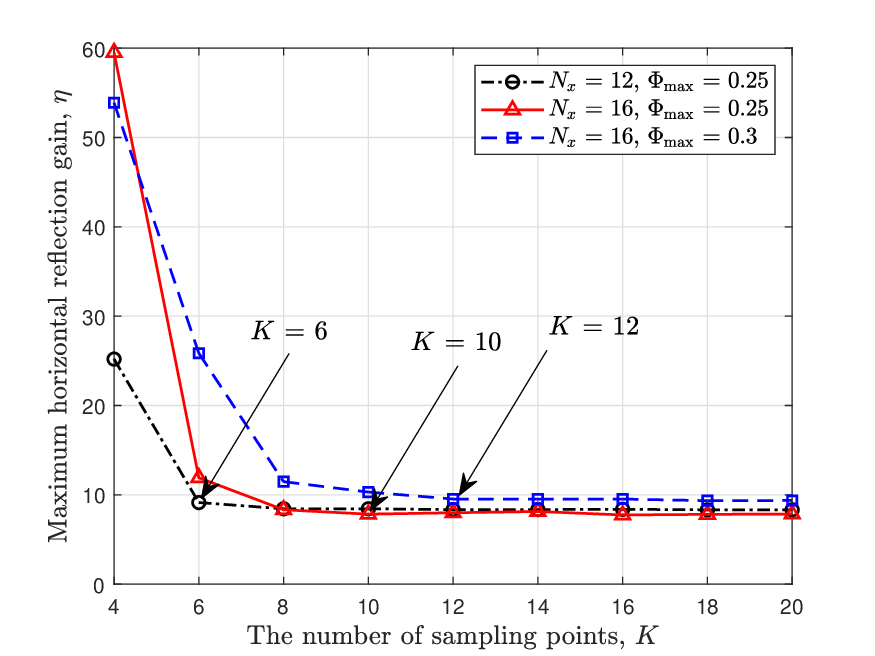}
		\caption{Maximum horizontal reflection gain obtained by the proposed IRS-aided ES system versus the number of sampling points $K$.}
		\label{fig_samp}
	\end{minipage} \vspace{-0.1cm}
\end{figure*}

In Fig.~\ref{fig_beam}, we compare the horizontal reflection gain obtained by different approaches versus the spatial frequency $\Phi$ with $N_x =$ 16. We see that the proposed Lagrange dual-based optimization achieves a significantly lower reflection gain compared to the baseline system without IRS and the random phase shift design for any value of $\Phi$ within interval $\left[\Phi_{\mathrm{min}}, \Phi_{\mathrm{max}} \right]$, corresponding to the unauthorized detection region. Although the baseline system with a single sampling point can fully eliminate the reflection gain at $\Phi = 0$, it fails to neutralize the radar probing signals across the entire unauthorized detection region. The above results validate that the IRS design derived in \eqref{deqn_ex9b} can effectively reduce the reflection gain of the target and then neutralize the echo signals in the direction of the radar. In addition, the reflection gains obtained by our proposed scheme are approximately constant within the interval $\left[\Phi_{\mathrm{min}}, \Phi_{\mathrm{max}} \right]$, demonstrating that the proposed IRS-aided ES system acts like a \textit{dynamic band-stop spatial filter}, shielding radar probing signals echoed back from the target and reflected toward the unauthorized detected region. We also observe that the reflection gain is significantly enhanced around $\Phi = -$0.35 and $\Phi =$ 0.35, indicating that our proposed approach transfers the reflected electromagnetic energy outside the unauthorized detection region.

Fig.~\ref{fig_gain} shows the maximum horizontal reflection gain within the interval $\left[\Phi_{\mathrm{min}}, \Phi_{\mathrm{max}} \right]$ obtained by the considered methods versus the number of IRS elements $N_x$. As expected, the maximum reflection gain of the baseline system without IRS remains constant regardless of $N_x$ since in this case, $\mathbf{u}_k^T \bm{\theta}=0, \forall k=1,2,\dots,K$. Since increasing the size of the IRS augments the degrees of freedom available for echo signal manipulation at the target-mounted IRS, the maximum reflection gain of the proposed approach decreases with $N_x$. Due to the inevitable gain fluctuation for the designed beam that results from the discrete approximation for the continuous spatial frequency deviation, the maximum reflection gain of the proposed approach levels off but cannot achieve full ES (i.e., $\eta = 0$). In contrast, the baseline system with a single sampling point and the benchmark approach with a randomly configured IRS both actually enhances the reflection gain, which has the undesired effect of increasing the target detection probability. The above results confirm the effectiveness of proper passive IRS reflection design to achieve ES for an unauthorized detection region.

Fig.~\ref{fig_samp} plots the maximum horizontal reflection gain within the interval $\left[\Phi_{\mathrm{min}}, \Phi_{\mathrm{max}} \right]$ obtained by the proposed IRS-aided ES system versus the number of sampling points. It is observed that as $K$ increases, the maximum gains for different $N_x$ and $\Phi_{\mathrm{max}}$ values substantially decrease and quickly approach a constant $\eta$. Since the beamwidth of the phased array is inversely proportional to the array aperture $N_x\triangle_e$, and since a wider unauthorized detection region requires more sampling points to cover it to consistently reduce the reflection gain, we see that a larger $N_x$ and $\Phi_{\mathrm{max}}$ require more sampling points to achieve the minimum $\eta$.

\section{Conclusion}
In this letter, we proposed an IRS-aided ES strategy to evade potential radar detection within an unauthorized detection region. To neutralize the radar probing signals echoed back from the target and reflected to any possible radar location in the unauthorized detection region, the IRS reflection pattern was designed to minimize the maximum received SNR across the area. In particular, the IRS reflection coefficients were optimized by discretizing the angular region to be nulled and then applying the Lagrange dual method. Simulation results validated the proposed ES strategy of placing a suitably designed IRS on the target surface to achieve satisfactory stealth performance for an unauthorized detection region.


\newpage

\vfill


\begin{thebibliography}{1}
\bibliographystyle{IEEEtran}
\vspace{-0.1cm}

\bibitem{ref_Stealth}
G. Rao and S. Mahulikar, ``Integrated review of stealth technology and its role in airpower,'' \textit{Aeronaut. J.}, vol. 106, no. 1066, pp. 629–642, Dec. 2002.

\bibitem{ref_Mater_ref}
B. Bai, X. Li, J. Xu, and Y. Liu, ``Reflections of electromagnetic waves
obliquely incident on a multilayer stealth structure with plasma and radar absorbing material,'' \textit{IEEE Trans. Plasma Sci.}, vol. 43, no. 8, pp. 2588–2597, Aug. 2015.

\bibitem{ref_Mater_abs}
D. Micheli, A. Vricella, R. Pastore, and M. Marchetti, ``Synthesis and
electromagnetic characterization of frequency selective radar absorbing
materials using carbon nanopowders,'' \textit{Carbon}, vol. 77, pp. 756–774, Jun. 2014.

\bibitem{ref_Wu_Tut}
Q. Wu, S. Zhang, B. Zheng, C. You, and R. Zhang, ``Intelligent reflecting surface aided wireless communications: A tutorial,'' \textit{IEEE Trans. Commun.}, vol. 69, no. 5, pp. 3313–3351, May 2021.

\bibitem{ref_Zheng_Sur}
B. Zheng, C. You, W. Mei, and R. Zhang, ``A survey on channel estimation and practical passive beamforming design for intelligent reflecting surface aided wireless communications,'' \textit{IEEE Commun. Surveys Tuts.}, vol. 24, no. 2, pp. 1035–1071, Second Quarter 2022.

\bibitem{ref_Zheng_TSP}
B. Zheng, X. Xiong, J. Tang, and R. Zhang, ``Intelligent reflecting surface-aided electromagnetic stealth against radar detection,'' \textit{IEEE Trans. Signal Process.}, vol. 72, pp. 3438–3452, Jun. 2024.

\bibitem{ref_Xiong_WCL}
X. Xiong, B. Zheng, A. L. Swindlehurst, J. Tang, and W. Wu, ``A new intelligent reflecting surface-aided electromagnetic stealth strategy,'' \textit{IEEE Wireless Commun. Lett.}, vol. 13, no. 5, pp. 1498–1502, May 2024.

\bibitem{ref_Zheng_Mag}
B. Zheng, X. Xiong, T. Ma, D. W. K. Ng, A. L. Swindlehurst, and R. Zhang, ``Intelligent reflecting surface-enabled anti-detection for secure sensing and communications,'' \textit{IEEE Wireless Commun.}, Early Access,
2024.

\bibitem{ref_RAmodeling}
B. Zheng, Q. Wu, and R. Zhang, ``Rotatable antenna enabled wireless communication: Modeling and optimization,'' \textit{arXiv preprint arXiv:2501.02595}, Jan. 2025.

\bibitem{ref_Luo_SDR}
Z. Luo, W. Ma, A. M. So, Y. Ye, and S. Zhang, ``Semidefinite relaxation
of quadratic optimization problems,'' \textit{IEEE Signal Process. Mag.}, vol. 27, no. 3, pp. 20–34, May 2010.

\bibitem{ref_IRShybrid}
F. Wang, H. Li, and A. L. Swindlehurst, ``Clutter suppression for target detection using hybrid reconfigurable intelligent surfaces,'' in \textit{Proc. IEEE Radar Conf. (RadarConf)}, San Antonio, TX, USA, May 2023, pp. 1–5.

\bibitem{ref_Boyd}
S. Boyd and L. Vandenberghe, \textit{Convex Optimization}. Cambridge, U.K.: Cambridge Univ. Press, Mar. 2004.

\bibitem{ref_Neunteufel_RCS}
D. Neunteufel, F. Galler, and H. Arthaber, ``Comprehensive measurement of complex-valued delta radar cross-section,'' in \textit{Proc. IEEE Int. EURASIP Workshop on RFID Technol. (EURFID)}, Brno, Czech Republic, Sep. 2018.

\end{thebibliography}
\end{document}